\documentclass[12pt,a4paper]{article}
\usepackage{a4wide}
\usepackage{latexsym}
\usepackage{epsf}
\usepackage{amssymb}
\begin{document}
\def\be{\begin{equation}}
\def\ee{\end{equation}}
\def\bea{\begin{eqnarray}}
\def\eea{\end{eqnarray}}

\def\pd{\partial}
\def\a{\alpha}
\def\b{\beta}
\def\g{\gamma}
\def\d{\delta}
\def\m{\mu}
\def\n{\nu}
\def \h{\mathcal{H}}
\def \hh{\mathcal{G}}
\def\t{\tau}
\def\p{\pi}
\def\th{\theta}
\def\l{\lambda}
\def\O{\Omega}
\def\r{\rho}
\def\s{\sigma}
\def\e{\epsilon}
  \def\scri{\mathcal{J}}
\def\cM{\mathcal{M}}
\def\tcM{\tilde{\mathcal{M}}}
\def\RR{\mathbb{R}}

\hyphenation{re-pa-ra-me-tri-za-tion}
\hyphenation{trans-for-ma-tions}


\begin{flushright}
IFT-UAM/CSIC-05-01\\
hep-th/0501146\\
\end{flushright}

\vspace{1cm}

\begin{center}

{\bf\Large   Can one tell Einstein's unimodular theory 
from Einstein's general relativity?}

\vspace{.5cm}

{\bf Enrique \'Alvarez }

\vspace{.3cm}

\vskip 0.4cm  
 
{\it  Instituto de F\'{\i}sica Te\'orica UAM/CSIC, C-XVI,
and  Departamento de F\'{\i}sica Te\'orica, C-XI,\\
  Universidad Aut\'onoma de Madrid 
  E-28049-Madrid, Spain }

\vskip 0.2cm

\vskip 1cm

{\bf Abstract}

\end{center}

\begin{quote}

The so called unimodular theory of gravitation is compared with general relativity
in the quadratic (Fierz-Pauli) regime, using a quite broad framework, 
 and it is argued that quantum effects
allow in principle to discriminate between both theories.  
\end{quote}


\newpage

\setcounter{page}{1}
\setcounter{footnote}{1}
\newpage
\section{Introduction}
Although it does not seem to be generally known (see, however, a footnote in 
\cite{Unruh}), four years after
writing down the equations of general relativity, Einstein \cite {Einstein} 
also proposed a different set of
equations, what have subsequently been dubbed as corresponding to 
{\em unimodular gravity}. The (english
translation of the) title of
Einstein's paper is: {\em Do gravitational fields play an essential part in the
structure of the elementary particles of matter?}, and its purpose was to obtain an alternative
to Mie's theory on the stability of the electron, and as such, it was unsuccessful. Quoting
Einstein himself: {\em \ldots Thus the problem of the constitution of the elementary 
quanta cannot 
yet be solved on the immediate basis of the given field equations}. 
\par
But on the
way, he realized already in 1919 that the unimodular theory is equivalent 
to general relativity, 
with the cosmological constant appearing as an integration constant. Let us quickly recall
 how this comes about.
\par
The posited equations of motion are the tracefree part of Einstein's general relativity
ones (written in dimension $n$):
\be\label{1919e}
R_{\m\n}-\frac{1}{n}R g_{\m\n}=\kappa^2 \left(T_{\m\n}-\frac{1}{n}T g_{\m\n}\right)
\ee
(with $\kappa^2\equiv 8\pi G$).
It seems that there is less information here, because the trace has been left out, but 
this is deceptive: the contracted Bianchi identities guarantee that
\be
\nabla^{\n}R_{\m\n}=\frac{1}{2}\nabla_{\m}R.
\ee
Applying this to the equations in the set (\ref{1919e}) we get (assuming covariant conservation 
of the energy-momentum tensor)
\be
\frac{n-2}{2}\nabla_{\m}R=-\frac{\kappa^2}{n}\nabla_{\m}T
\ee
which integrates to
\be
\frac{n-2}{2 d}R+\frac{2\kappa^2}{d}T = constant\equiv -\lambda
\ee
and plugging this fact into the equation (\ref{1919}) 
yields precisely Einstein's general relativity 
equations, with arbitrary cosmological constant, $\lambda$:
\be
R_{\m\n}-\frac{1}{2}(R +2 \lambda) g_{\m\n}=\kappa^2 T_{\m\n}
\ee

(Signs are chosen in such a way that with the Landau-Lifshitz spacelike conventions de Sitter
space corresponds to a positive cosmological constant).
\par
This result remains true when including higher order in curvature corrections, in the
following sense.
\par 
If the equations of motion can be derived from a covariant 
 action principle, schematically
\be
S\equiv \int d(vol) L(g_{\m\n},R_{\m\n\rho\sigma})
\ee 
there is a generalization of the Bianchi identity which ensures that
\be
\nabla_{\m}\frac{\d S}{\d g_{\m\n}}\equiv 0
\ee
This Bianchi identity, in turn, allows the trace $g_{\a\b}\frac{\d S}{\d g_{\a\b}}$ 
to be recovered as an integration constant
from the tracefree component of the equations of motion.
\par
Some previous work on this theory is \cite{Daughton}\cite{Henneaux}\cite{Kreuzer}
\cite{Ng}\cite{Unruh}\cite{vanderBij}\cite{Sato}\cite{Padma}.
\par
In the paper by van der Bij, van Dam and Ng, in particular, it is 
proven that the famous argument, coming from 
the need to represent trivially the translations of the little group of massless particles, 
$ISO(2)$
that leads to abelian $U(1)$ gauge invariance in the spin one case, 
also leads to the unimodular theory in the spin two case. This means that it is enough
for this purpose (namely, to represent translations trivially), to impose the gauge
symmetry at the linear level
\be
\d h_{\a\b}=\pd_{\a}\xi_{\b}+\pd_{\b}\xi_{\a}
\ee
with 
\be
\pd_{\a}\xi^{\a}=0
\ee
instead of the Fierz-Pauli \cite{Fierz} full symmetry without the transversality condition.
\par
The purpose of the present work is the quite modest one of answering the question in the title,
as to whether it is at all possible to discrimitate between the two Einstein's theories.
Al the classical level, the equations are identical, so it seems that attention
should be focused in quantum effects.

\section{The action principle}
Einstein never talks about an action principle in his paper. And this for a good reason,
given the fact that
he was only interested in the equations of motion. But if we are interested in quantum
effects, this is not enough, and we have to invoke a particular lagrangian. There are several
ways to do this, and each one of them defines {\em a priori} a different quantum theory.
Some of the alternatives have been discussed in the references, in particular in \cite{Unruh}
and in \cite{Henneaux}.
\par
The full action comprising the metric and the matter fields, $\phi_i$, will be represented as
\be
S=S_{grav}+ S_{matt}
\ee
with
\be
S_{grav}\equiv \int \sqrt{|g|}d^n x L(g_{\a\b},R_{\a\b\gamma\d})
\ee
and
\be
S_{matt}\equiv \int \sqrt{|g|}d^n x L(g_{\a\b},\phi_i)
\ee

The allowed variations , which we be denoted as $\d^t g_{\a\b}$ are constrained by
\be
\d^t \sqrt{|g|}=0
\ee
that is
\be\label{t}
g^{\a\b}\d^t g_{\a\b}=0
\ee
which can be expressed in terms of an unconstrained variation as
\be
\d^t g_{\a\b}=\d g_{\a\b}-\frac{1}{n}g_{\a\b} g^{\m\n}\d g_{\m\n}
\ee
It is clear that any variation can be expressed as
\be
\d S = \int d^n x \frac{\d S}{\d g_{\a\b}}\d g_{\a\b}=
\int d^n x \frac{\d S}{\d g_{\a\b}}\left(\d^t g_{\a\b}+\frac{1}{n}g_{\a\b} g^{\m\n}
\d g_{\m\n}\right)
\ee
so that the restricted variation is just the trace-free part of the unconstrained variation:
\be
\frac{\d S}{\d^t g_{\a\b}}=\frac{\d S}{\d g_{\a\b}}-\frac{1}{n}g^{\m\n}\frac{\d S}{\d g_{\m\n}}
g_{\a\b}
\ee
and this variational principle indeed yields Einstein's unimodular field equations as written
down in the introduction when the Hilbert lagrangian is used as an starting point.
\par

Were we to put forward a stronger claim, namely that all physics 
is invariant under APD only, which would then the basic symmetry principle which takes the place
of diffeomorphism invariance (DI), then the energy momentum tensor would not be fully 
covariantly conserved.
The assumed symmetry only guarantees that
\be
\int\sqrt{|g|}d^n x (\nabla_{\a}\xi_{\b}+\nabla_{\b}\xi_{\a})\frac{\d L_{matt}}{\d g_{\a\b}}=0
\ee
whenever the vector field $\xi$ satisfies the transversality condition
\be
\nabla_{\a}\xi^{\a}=0
\ee
which can be locally integrated in terms of an $(n-2)$-differential form, $\Omega$,  to
\be
\xi^{\a}=\frac{1}{\sqrt{|g|}}\d^{\a\b\m_1\ldots\m_{n-2}}\nabla_{\b}\Omega_{\m_1\ldots\m_{n-2}}
\ee
The conservation law for $\Theta_{\a\b}\equiv \frac{2}{\sqrt{|g|}}\frac{\d S_{matt}}
{\d g_{\a\b}}$ 
is then

\bea
&&\int\sqrt{|g|}d^n x (\nabla^{\a}\frac{1}{\sqrt{|g|}}\d^{\b\d\m_1\ldots\m_{n-2}}\nabla_{\d}
\Omega_{\m_1\ldots\m_{n-2}}+\nonumber\\
&&\nabla^{\b}\frac{1}{\sqrt{|g|}}\d^{\a\d\m_1\ldots\m_{n-2}}\nabla_{\d}
\Omega_{\m_1\ldots\m_{n-2}})\Theta_{\a\b}
\eea

that is
\be
\nabla_{\m}\Theta^{\m\n}=\nabla^{\n} \phi
\ee
for some scalar $\phi$. 

This is consistent with our previous finding that
\be
\Theta_{\a\b}=T_{\a\b}-\frac{1}{n}T g_{\a\b}
\ee
in such a way that indeed
\be
\nabla_{\a}\Theta^{\a\b}=\frac{1}{n}\nabla^{\b}T
\ee

\section{Quadratic analysis}
Let us start with the well-known analysis which leads eventually to the
Fierz-Pauli lagrangian for a free massless spin two particle (cf. \cite{Veltman},\cite{Ortin}).
 A simple road is as follows: 
the quadratic part of the 
lagrangian is the inverse of the propagator, and the propagator is related to the
possible polarizations. 
There are five of those in the massive spin two case, which can be represented as $\e_{\m\n}^A$
$A=1\ldots 5$, with

\bea
&&\e^A_{\m\n}=\e^A_{\n\m}\nonumber\\
&&k^{\m}\e^A_{\m\n}=0\nonumber\\
&&\eta^{\m\n}\e^A_{\m\n}=0
\eea
We can expand the momentum space \footnote{Both position and momentum space notation
will be used for convenience. Although most formulas will be written in arbitrary dimension,
most of the polarization reasoning is implicitly four-dimensional.} propagator in terms 
of the basic tensors  
$k^{\m}$ and the off-shell
transverse projection operator $\eta^T_{\m\n}\equiv
\eta_{\m\n}-\frac{k_{\m}k_{\n}}{k^2}$ as
\bea
&&D_{\m\n\lambda\sigma}\equiv\sum_A\e_{\m\n}^A\e^A_{\lambda\sigma}= 
c_1 \eta^T_{\m\n}\eta^T_{\lambda\sigma}+ c_2 \eta^T_{\m\n}k_{\lambda}k_{\sigma}+ 
k_{\m} k_{\n} \eta^T_{\lambda\sigma}\nonumber\\
&&+c_3 (\eta^T_{\m\lambda}\eta^T_{\n\sigma} + \eta^T_{\m\sigma} \eta^T_{\n\lambda})
+c_4(k_{\m}k_{\sigma}\eta^T_{\n\lambda}+k_{\m}k_{\lambda}\eta^T_{\n\sigma}+\nonumber\\
&&k_{\n}k_{\sigma}\eta^T_{\m\lambda}+k_{\n}k_{\lambda}\eta^T_{\m\sigma}+
c_5 k_{\m}k_{\n}k_{\lambda}k_{\sigma}
\eea
Imposing off-shell transversality and tracelessness we get uniquely
\be\label{propa}
D_{\m\n\lambda\sigma}= c_1\left(\eta^T_{\m\n}\eta^T_{\lambda\sigma}-\frac{3}{2}
(\eta^T_{\m\lambda}\eta^T_{\n\sigma}+\eta^T_{\m\sigma}\eta^T_{\n\lambda})\right)
\ee

Acting on conserved currents, we can drop the superscript $T$.
\par
In order to find the lagrangian, we have to compute the propagator by imposing
 transversality on shell only. Otherwise
there are unwanted degeneracies. This amounts to change the projector in (\ref{propa}) 
$\eta_{\m\n}^T$ for  a quantity
$\eta_{\m\n}^{TOS}\equiv\eta_{\m\n}-\frac{k_{\m}k_{\n}}{m^2}$ , which
 behaves as a projector on shell only:
\bea
&&\eta_{\m\n}^{TOS}k^{\n}=k_{\m}\frac{m^2-k^2}{m^2}\nonumber\\
&&\eta^{TOS}_{\m\n}\eta^{\m\n}=3+\frac{m^2-k^2}{m^2}\nonumber\\
&&\eta^{TOS}_{\m\n}(\eta^{TOS})^{\n\rho}=\eta^{TOS}_{\m}\,^{\rho}+\frac{k^2-m^2}{m^4}
k_{\m}k^{\rho}
\eea
What remains is
\be\label{ftm}
D^m_{\m\n\lambda\sigma}= c_1\left(\eta^{TOS}_{\m\n}\eta^{TOS}_{\lambda\sigma}-\frac{3}{2}
(\eta^{TOS}_{\m\lambda}\eta^{TOS}_{\n\sigma}+\eta^{TOS}_{\m\sigma}\eta^{TOS}_{\n\lambda})\right)
\ee
The lagrangian is then found by computing the inverse.
\be
(K^m)_{\m\n\a\b} (D^m)^{\a\b}\,_{\lambda\d} = \frac{1}{2}(\eta_{\m\lambda}\eta_{\n\d}+ 
\eta_{\m\d}
\eta_{\lambda\n})
\ee
The conventional normalization corresponds to
\be
c_1=-\frac{4}{3}\frac{1}{k^2-m^2}
\ee
and yields
\bea
&&(K^m)_{\m\n\rho\sigma}=\frac{k^2-m^2}{8}(\eta_{\m\rho}\eta_{\n\sigma}+\eta_{\m\sigma}
\eta_{\n\rho}-2\eta_{\m\n}\eta_{\rho\sigma})\nonumber\\
&&-\frac{1}{8}(k_{\m}k_{\rho}\eta_{\n\sigma}+k_{\n}k_{\sigma}\eta_{\m\rho}+
k_{\m}k_{\sigma}\eta_{\n\rho}+k_{\n}k_{\rho}\eta_{\m\sigma}-2k_{\m}k_{\n}\eta_{\rho\sigma}
-2k_{\rho}k_{\sigma}\eta_{\m\n})
\eea
which corresponds to the Fierz-Pauli lagrangian
\be\label{fierzpauli}
L_{FP}=\frac{1}{4}\pd_{\m}h^{\n\rho}\pd^{\m}h_{\n\rho}-
\frac{1}{2}\pd_{\m}h^{\n\rho}\pd^{\n}h_{\m\rho}+
\frac{1}{2}\pd_{\m}h\pd^{\s}h_{\s\m}-
\frac{1}{4}\pd_{\m}h\pd^{\m}h-\frac{m^2}{4}(h_{\a\b}h^{\a\b}-h^2)
\ee
where $h\equiv \eta^{\m\n}h_{\m\n}$.
\par
It follows that
\be\label{uno}
k^{\n}K^m_{\m\n\rho\sigma}h^{\rho\sigma}=-2 m^2(k^{\rho}h_{\rho\m}- k_{\m} h)
\ee
so that necessarily,
\be
k^2 h = k_{\rho}k_{\sigma}h^{\rho\sigma}
\ee
The trace gives:
\be
\eta^{\m\n}K^m_{\m\n\rho\sigma}h^{\rho\sigma}=-2(1-n)m^2 h
\ee
which in turn implies that
\be
h=k_{\m}k_{\n}h^{\m\n}=0
\ee
and using (\ref{uno}),
\be
k^{\m}h_{\m\n}=0
\ee
so that the field obeys the Klein-Gordon equation
\be
(\Box+m^2)h_{\m\n}=0
\ee
\par
It can be shown (\cite{van}) that this particular mass term is the only one which is compatible 
with unitarity.
\par
\subsection{The massless limit.}
The massless limit is singular.
Three polarizations can be written as
\be
k_{\m}u_{\n}+k_{\n}u_{\m}
\ee
with $k.u=0$. Namely, in an obvious notation, ($e_{(a)}\equiv \pd_a$, etc)
\bea
&&k\otimes k\nonumber\\
&&k\otimes e_{(1)}+e_{(1)}\otimes k\nonumber\\
&&k\otimes e_{(2)}+e_{(2)}\otimes k
\eea
The remaining two are
\bea\label{pola}
&&\e_1\equiv e_{(1)}\otimes e_{(2)}+e_{(2)}\otimes e_{(1)}\nonumber\\
&&\e_2\equiv e_{(1)}\otimes e_{(1)}-e_{(2)}\otimes e_{(2)}\nonumber\\
\eea
and under the little group, they transform into the other three (cf.\cite{vanderBij}).
\par
This means that exactly the same type of reasoning that gives rise to the abelian 
gauge invariance yields the unimodular theory of Einstein, which is invariant under
area preserving diffs only:
\be
\d h_{\m\n}=\pd_{\m}\xi_{\n}+\pd_{\n}\xi_{\m}
\ee
with
\be\label{area}
\pd_{\m}\xi^{\m}=0
\ee

Once we implement this symmetry (with or without the unimodularity condition (\ref{area})),
 then there is a gauge in which the massless Fierz-Pauli 
propagator is defined up to a constant as:
\be
D^{GF}_{\m\n\r\s}=c_2(\eta_{\m\r}\eta_{\n\s}+\eta_{\m\s}\eta_{\n\r}-\eta_{\m\n}\eta_{\r\s})
\ee
And then, it is a simple matter to show that, acting on conserved currents,
\be
D^{GF}_{\m\n\r\s}=D^m_{\m\n\r\s}+\frac{c_1}{2}\eta_{\m\n}\eta_{\r\s}
\ee
which means that there is an extra admixture of spin $s=0$ in the massless case.
\par
The conventional normalization corresponds to
\be
c_2=\frac{4}{k^2}
\ee
and yields
\be
K^{GF}_{\m\n\r\s}=\frac{k^2}{8}(\eta_{\m\r}\eta_{\n\s}+\eta_{\m\s}\eta_{\n\r}-
\eta_{\m\n}\eta_{\r\s})
\ee
This corresponds to the massless Fierz-Pauli lagrangian with the harmonic gauge
condition
\be
L_{GF}=\frac{1}{2}(\pd_{\n}h_{\m}\,^{\n}-\frac{1}{2}\pd_{\m}h)^2
\ee
that is
\be
L_0= \frac{1}{4}(\pd_{\m}h_{\a\b})^2-\frac{1}{8}(\pd_{\m}h)^2
\ee
\subsection{Unimodular lagrangians}

If we implement the restricted gauge symmetry only, a simpler lagrangian exists:
\be\label{1919}
L_{u}= \frac{1}{4}(\pd_{\m}h_{\a\b})^2- \frac{1}{2}\pd_{\m}h_{\a\b}\pd^{\a}h^{\m\b}
\ee
although the full Fierz-Pauli lagrangian $L_{FP}$ is obviously still invariant under
the restricted symmetry.
This is exactly the same thing that would have been gotten by putting $h=0$ 
in the Fierz-Pauli lagrangian, that is
\bea
&&(K_u)_{\m\n\rho\sigma}=\frac{k^2}{8}(\eta_{\m\rho}\eta_{\n\sigma}+
\eta_{\m\sigma}\eta_{\n\rho})+\nonumber\\
&&-\frac{1}{8}( k_{\m}k_{\rho}\eta_{\n\sigma}+k_{\rho}k_{\n}\eta_{\m\sigma}+
 k_{\sigma}k_{\n}\eta_{\m\rho}+k_{\sigma}k_{\m}\eta_{\n\rho})
\eea
\par
Let us now construct a massive unimodular theory. In order to do that, we postulate the 
most general mass term, say
\be\label{1919m}
-\frac{m^2}{8}(2 h_{\m\n}h^{\m\n}-r h^2)
\ee
where $r$ is an arbitrary constant (which for the full Fierz-Pauli theory happens to take
the value  $r=2$). The posited full kinetic operator is then
\bea
&&(K_u^m)_{\m\n\rho\sigma}=\frac{k^2-m^2}{8}(\eta_{\m\rho}\eta_{\n\sigma}+
\eta_{\m\sigma}\eta_{\n\rho})+\nonumber\\
&&r\frac{m^2}{8}\eta_{\m\n}\eta_{\rho\sigma}
-\frac{1}{8}(k_{\m}k_{\rho}\eta_{\n\sigma}+k_{\rho}k_{\n}\eta_{\m\sigma}+
k_{\sigma}k_{\n}\eta_{\m\rho}+k_{\sigma}k_{\m}\eta_{\n\rho})
\eea
The corresponding equation of motion is:
\be
(K^m_u h)_{\m\n}=\frac{k^2-m^2}{4}h_{\m\n}+\frac{r m^2}{8}h \eta_{\m\n}-\frac{1}{4}(k_{\m}
k_{\rho}h_{\n}\,^{\rho}+k_{\n}k_{\rho}h_{\m}\,^{\rho})
\ee

Computing again the transverse part of the equation of motion:

\be
(K^m_u.h)_{\m\n}=\frac{k^2-m^2}{4}h_{\m\n}+ r \frac{m^2}{8} h \eta_{\m\n}-\frac{1}{4}(k_{\m}
k^{\rho}h_{\n\rho}+k_{\n}k^{\rho}h_{\m\rho})
\ee

\be
k^{\m}k^{\n}(K^m_u)_{\m\n\rho\sigma}h^{\rho\sigma}=
-(k^2+m^2)k_{\rho}k_{\sigma}h^{\rho\sigma}+ r\frac{
m^2 k^2}{2}h=0
\ee
and the trace:
\be 
\eta^{\m\n}(K^m_u)_{\m\n\rho\sigma}h^{\rho\sigma}=(k^2 -m^2+\frac{n}{2} r m^2)h - 2 
k_{\rho}k_{\sigma}h^{\rho\sigma}=0
\ee
This two conditions enforce
\be
h=k_{\rho}k_{\sigma}h^{\rho\sigma}=0
\ee
as long as $r>0$. Even when $r=0$ they do enforce
full transversality, although tracelessness is then only guaranteed off shell
\be
(k^2-m^2)h=0
\ee
The conclusion of this analysis is that the unimodular theory becomes massive with a mass term
of the Fierz-Pauli type.

\subsection{Propagators}
Logically, our attention should now turn to a discussion of the unimodular massive propagator.
The fact is that, for the minimal model (\ref{1919}), supplemented by a mass term such as
the one in [\ref{1919m}], there is no propagator, because this lagrangian is singular.
This is perhaps somewhat of a surprise, because there is no known gauge symmetry when the mass
is nonvanishing, but it is nevertheless true. Actually, the situation is as follows: 
there is a particular mode,
proportional to
\be
(\eta^T_u)_{\rho\sigma}\equiv (k^2 +m^2 -r \frac{m^2}{2})\eta_{\rho\sigma}-(k^2+m^2-r \frac{n}
{2}m^2) \frac{k_{\rho}k_{\sigma}}{k^2}
\ee
such that
\be
(K^m_u \eta^T_u)_{\m\n}
\ee
is transverse, i.e.
\be
(K^m_u \eta^T_u)_{\m\n}k^{\m}=0
\ee
Although this is not a zero mode {\em sensu stricto}, it is enough to make the 
lagrangian singular.
The situation is somewhat strange. Nevertheless, we already know, because of the 
argument of the polarizations at the beginning
of the present section, that the correct lagrangian for  massive spin 2 is the Fierz-Pauli one,
(\ref{fierzpauli}). On the other hand, we know that the model (\ref{1919} \ref{1919m}) is the
 minimal one which can be extended to {\em exactly} the Fierz-Pauli one while keeping only the
restricted gauge symmetry in the massless case.
\par
While it would be interesting to further study the minimal theory, we shall therefore confine
 our attention from now on to the Fierz-Pauli lagrangian.

\par
\subsection{Gauge fixing}
The harmonic gauge is not reachable in the massless unimodular  theory, because the equation
\be
\pd_{\m}h_{(0)}^{\m\n}-\frac{1}{2}\pd^{\n}h_{(0)}+\Box\xi^{\n}=0
\ee
has got the integrability condition
\be
2 \pd_{\a}\pd_{\b}h_{(0)}^{\a\b}=\Box h_{(0)}
\ee
Incidentally, the same holds true for any covariant linear gauge which is 
also linear in derivatives, that
is, of the form:
\be
M_{\a\b\gamma}h^{\b\gamma}=0
\ee
with
\be
M_{\a\b\gamma}=c_1 \eta_{\a\b}k_{\gamma}+c_2 \eta_{\a\gamma}k_{\b}+ c_3 \eta_{\b\gamma}k_{\a}
\ee
\par
The equations of motion before gauge fixing are
\be
\Box h_{\a\b}=\pd_{\b}\pd^{\s}h_{\a\s}+\pd_{\a}\pd^{\s}h_{\b\s}
\ee
which lead to
\be\label{prp}
\Box h=2 \pd_{\a}\pd_{\b}h^{\a\b}
\ee
so that on shell the equations of motion are equivalent to the Fierz-Pauli ones, and the
harmonic gauge is possible. This is true in spite of the fact that the unimodular wave operator
is neither transverse ($k^{\m}(K_u)_{\m\n\rho\sigma}\neq 0$) nor traceless 
($\eta^{\m\n}(K_u)_{\m\n\rho\sigma}\neq 0$).
\par
Indeed the two extra terms that appear in the Fierz-Pauli 
equations of motion are
\be
-\frac{1}{2}\eta_{\m\n}\pd_{\a}\pd_{\b}h^{\a\b}+\frac{1}{4}\eta_{\m\n}\Box h
\ee
which vanish due to (\ref{prp}), which still holds here.
We knew already that much, because we have proven the on-shell equivalence of the two theories 
at the nonlinear level in the introduction.
\par
When we want to define the quantum theory, by means of a path integral, for example, we could
impose in the Fierz-Pauli theory the noncovariant gauge conditions
\bea
&&h=0\nonumber\\
&&\pd_{\m}h^{\m i}=0
\eea
whereas in the minimal unimodular theory we could impose
\be
\pd_{\m}h^{\m i}=0
\ee
(we can consider that only the $\xi^i$ are independent, whereas $\xi^0=\int^{x^0}
\vec{\nabla}\vec{\xi}$.)
But the trace $h$ is not necessarily zero, so that the action off shell differs in general 
from the
Fierz-Pauli value, and it is easy to convince oneself that there is not a gauge in which the
two actions coincide. This is enough to show that the two theories are different at the 
quantum level, because the action measures the quantum phase associated to each path. 
\par
All this remains true if we choose the full Fierz-Pauli lagrangian $L_{FP}$ as our
starting point before restricted gauge fixing. In that case, the unimodular path integral is
\be
Z_u(J)=\int {\cal D}h \, e^{i S_{FP} + i \int d^n x \frac{1}{2\a}(\pd_{\m}h^{\m i})^2 + 
i\int d^n x J_{\a\b}h^{\a\b}}
\ee
whereas with the full symmetry the closest we can get is
\be
Z(J)=\int {\cal D}h\,  \d(h)\, e^{i S_{FP} + i \int d^n x \frac{1}{2\a}(\pd_{\m}h^{\m i})^2+
i\int d^n x J_{\a\b}h^{\a\b}}
\ee
We are not worrying about ghosts because we remain in the abelian approximation.
The conclusion of the present analysis is that even in the non-interacting case the two
theories are different, and this in spite of the fact that the starting lagrangian is 
the same, owing
to the different gauge fixing.

\section{Full covariant unimodular lagrangians.}

One of the most interesting full covariant lagrangian (i.e., diffeomorphism invariant) 
versions of the same traceless equations
of motion is the one proposed in \cite{Henneaux}, in terms of a ($n-1$)-differential form, 
$A_{n-1}$ and
a scalar field, $\lambda(x)$, namely
\be
S=\frac{1}{2\kappa^2}\left(\int d^n x \sqrt{|g|} (R+2\lambda)-2\int A\wedge d\lambda\right)
\ee
The equations of motion for the metric tensor are Einstein's equations with a cosmological
scalar function:
\be
\frac{\d S}{\d g^{\a\b}}=R_{\a\b}-\frac{1}{2}(R+2 \lambda(x))g_{\a\b}
\ee
The equation of motion for the field $\lambda(x)$ is
\be
\frac{\d S}{\d \lambda}=\sqrt{|g|}d^n x-d A
\ee
and finally, the equation of motion for the $A$-form is:
\be
\frac{\d S}{\d A}=d\lambda
\ee
There is an extra symmetry of the theory, namely
\be
\d A= d\Omega
\ee
which is the one that plays the r\^ole of the unimodular transformation, just by
defining
\be
\e^{\m}\equiv \frac{1}{2}\d^{\m\a_1\ldots\a_{n-1}}\pd_{\a_1}\Omega_{a_2\ldots \a_{n-1}}
\ee
which indeed obeys
\be
\pd_{\m}\e^{\m}=0
\ee

In this case it is even more clear than in the examples above that the full quantum theory
is different in principle from the one stemming from Einstein-Hilbert's lagrangian, at least
insofar as both of them possess different action principles.

\section{Conclusions}
It is well-known that the coupling of matter to the Fierz-Pauli lagrangian is inconsistent
unless the full nonlinear general relativity is reconstructed (cf. for
example, the general review in \cite{Alvarez}, where further references can be found).
From this point of view, it seems that the natural nonlinear completion of the unimodular
theory is the Einstein-Hilbert lagrangian, dressed with arbitrary functions of the determinant
of the metric.
\par
A straightforward calculation would then predict different perturbative amplitudes for
both theories. Outside the realm of perturbation theory, all speculation is possible.
An interesting question is, for example, how this restricted unimodular symmetry can
(if at all) be implemented in a scheme such as the one called {\em loop quantum gravity}.


\par
Given the fact that the equations of motion are identical, and, as we have shown, this property
remains true when higher order (in the curvature) corrections are considered,
it is clear that the identification of the low energy limit of string theories as 
general relativity is premature; it could easily be the unimodular theory we are considering. 
Reliable computation of stringy off-shell correlators could, of course, be decisive 
in this respect.

\par


Let us consider (cf. for example, \cite{Zee}) the free energy in the presence of
arbitrary conserved sources. This quantity is an exceedingly useful one to consider,
because in summarizes in a very simple way the physical content of the theory.
We shall assume two spatially disconnected sources:
 $T_{\a\b}\equiv(T_1)_{\a\b}\d^{(3)}(\vec{x}-\vec{x}_1)+
(T_2)_{\a\b}\d^{(3)}(\vec{x}-\vec{x}_2)$, with
\be
\pd_{\a}(T_1)^{\a\b}=\pd_{\a}(T_2)^{\a\b}=0
\ee
Keeping only the term bilinear in the sources, assumed to act for a total time interval
$\int dx^0\equiv T$, one easily gets:
\be
W=- \frac{2}{3}T \int d^3 k\frac{1}{\vec{k}^2+m^2}e^{i\vec{k}(\vec{x}-\vec{y})}E_{12}
\ee 
Starting with the massive Fierz-Pauli theory, the answer stemming from (\ref{ftm}) is
\be
E_{12}=\left( tr\,T_1 tr\,T_2 -3 tr\, T_1 T_2\right)
\ee

In the massless case, the Fierz-Pauli interacion energy in the harmonic gauge 
is  proportional instead to
\be
E_{12}\equiv \frac{1}{2}\left(2 tr T_1 T_2-(tr\,T_{1})(tr\,T_{2})\right)
\ee
Even forgetting about the coefficients, there is a mismatch of $3/2$ in the term $tr\,T_1 T_2$;
this is the famous van Dam-Veltman discontinuity (\cite{vanDam}), which indicates that there
is some sort of non smoothness in the massless limit.
\par
In full 
\footnote{t
The resulting expression can be further simplified  using current conservation:
\bea
&&T^{00}=\frac{\kappa}{\omega}T^{03}=\frac{\kappa^2}{\omega^2}T^{33}\nonumber\\
&&T^{0i}=\frac{\kappa}{\omega}T^{3i}
\eea
getting
\bea
E_{12}=&&\frac{1}{2}(T_1^{11}-T_1^{22})(T_2^{11}-T_2^{22})+ 2 T_1^{12}T_2^{12}\nonumber\\
&&+\frac{m^2}{2\omega^2}(T_1^{11}+T_1^{22})T_2^{33}+2\frac{m^2}{\omega^2}(T_1^{13}T_2^{13}
+T_1^{23}T_2^{23})\nonumber\\
&&+\frac{1}{2}T_1^{33}\left(\frac{m^4}{\omega^4}T_2^{33}-\frac{m^2}{\omega^2}(T_2^{11}+T_2^{22})
\right)
\eea
This clearly shows that in the massless limit only the two polarizations in (\ref{pola})
contribute (cf. \cite{Zee}\cite{Dicus}) to this physical observable.}
detail:
\bea\label{naive}
E_{12}=&&\frac{1}{2}T_1^{00}(T_2^{00}+T_2^{11}+T_2^{22}+T_2^{33})+
\frac{1}{2}T_1^{11}(T_2^{00}+T_2^{11}-T_2^{22}-T_2^{33})+\nonumber\\
&&\frac{1}{2}T_1^{22}(T_2^{00}-T_2^{11}+T_2^{22}-T_2^{33})+
\frac{1}{2}T_1^{33}(T_2^{00}-T_2^{11}-T_2^{22}+T_2^{33})+\nonumber\\
&&2\left(T_1^{12}T_2^{12}+T_1^{13}T_2^{13}+T_1^{23}T_2^{23}-
T_1^{01}T_2^{01}-T_1^{02}T_2^{02}-T_1^{03}T_2^{03}\right)
\eea

In order to identify possible off-shell intermediate states in the massless
Fierz-Pauli theory,
it is useful to transform the expression (\ref{naive}) (that is, before using 
current conservation) 
into the suggestive form proposed by
Dicus and Willenbrock \cite{Dicus}.
\bea\label{dicus}
E_{12}=&&\frac{1}{2}(T_1^{11}-T_1^{22})(T_2^{11}-T_2^{22})+ 2 T_1^{12}T_2^{12}+\nonumber\\
&&2\left(T_1^{13}T_2^{13}+T_1^{23}T_2^{23}-T_1^{01}T_2^{01}-T_1^{02}T_2^{02}\right)*\nonumber\\
&&\frac{1}{6}[2(T_1^{00}-T_1^{33})+T_1^{11}+T_1^{22}][2(T_2^{00}-T_2^{33})+T_2^{11}+T_2^{22}]
\nonumber\\
&&-\frac{1}{6}[-T_1^{00}+T_1^{11}+T_1^{22}+T_1^{33}][-T_2^{00}+T_2^{11}+T_2^{22}+T_2^{33}]
\nonumber\\
\eea
This can be easily checked: in order for the coefficient of $T_1^{00}$ in (\ref{dicus}) to be
the same as the one in (\ref{naive}) we have to add a term $T_1^{00}T_2^{33}$, and also
if we want the coefficient of $T_1^{33}$ in (\ref{dicus}) to be the same as in (\ref{naive})
we have to add another term $T_1^{33}T_2^{00}$. But in order for the coefficients of $T_1^{03}$
to match, we have to add $-2 T_1^{03}T_2^{03}$, which exactly cancel owing to the conservation
of the energy momentum tensor.

\par
This expansion can be spelled down physically as follows. Let us introduce a real basis of 
polarizations
in the generic case as
\bea
&&\e_3= e_{(0)}\otimes e_{(0)}-e_{(1)}\otimes e_{(1)}\nonumber\\
&&\e_4= e_{(0)}\otimes e_{(1)}+e_{(1)}\otimes e_{(0)}\nonumber\\
&&\e_5= e_{(0)}\otimes e_{(2)}+e_{(2)}\otimes e_{(0)}
\eea
Then the second line of (\ref{dicus}) is proportional to:
\be
T_1^{\m\n}(2\e_4+\e_5)_{\m\n}(2\e_4+\e_5)_{\r\s}T_2^{\r\s}
\ee
and the third one to
\be
T_1^{\m\n}(\e_2 + 2 e_3)_{\m\n}(\e_2 + 2 e_3)_{\r\s}T_2^{\r\s}
\ee
whereas the last one is a spin zero contribution
\be
T_1^{\m\n}(\e_1 + \e_2 +  e_3)_{\m\n}(\e_1+ \e_2 +  e_3)_{\r\s}T_2^{\r\s}
\ee
which obviously does not correspond to spin two, but is nevertheless neccessary to cancel the 
contribution of the unphysical polarizations in the massless case. So that not only are
off-shell spin zero components allowed by the theory as intermediate states,
but as has been pointed out by Dicus and Willenbrok, they are actually neccessary for 
consistency. The appearance of these components was first pointed out in \cite{Datta}.

\par
Coming back to our main theme, a natural question is how can we experimentally 
discriminate between both theories? There is an easy answer, namely that graviton scattering 
amplitudes 
are expected to be different in detail. But unfortunately, graviton scattering data do not
abound.
\par
A most interesting, and perhaps feasible 
experiment would be to weigh the vacuum energy, i.e. Casimir energy. Indeed, under the 
restricted variations in (\ref{t}) which we have labelled $\d^t g_{\m\n}$, the vacuum energy
does not affect \footnote{This point has been developed in discussions with Tom\'as Ort\'{\i}n.}
 the equations of motion. 

\par
A related point is the following. Granting that the two Einstein theories are 
indeed different at the 
quantum level, the most important physical question is 
whether this improves or otherwise reformulates 
in some way the problem of the cosmological constant.
Interesting suggestions in this direction have been made by \cite{Tseytlin} and 
\cite{Arkani-Hamed}, although no compelling model exists yet.

\section*{Acknowledgments}
I have benefited from
discussions with E. Gabrielli, Tom\'as Ort\'{\i}n and (long time ago, but still relevant 
for this paper) 
with Tini Veltman.
This work has been partially supported by the
European Commission (HPRN-CT-200-00148) and by FPA2003-04597 (DGI del MCyT, Spain).


\end{document}